\documentclass{elsart}
\usepackage{natbib}
\begin{document}
\runauthor{Shchur}
\begin{frontmatter}
\title{On the quality of random number generators with taps}
\author{Lev N. Shchur\thanksref{Support}\thanksref{email}}
\address{Landau Institute for Theoretical Physics, 142432
Chernogolovka, Russia}
\thanks[Support]{Partially supported by NWO 07-13-210, RFBR 96-02-18618
and 96-07-89266, INTAS 93-211 and Italian Ministry for Foreign Affairs}
\thanks[email]{lev@landau.ac.ru}
\begin{abstract}

Recent exact analytical results developed for the random number generators
with taps are reported. These results are applicable to a wide class of
algorithms, including random walks, cluster algorithms, Ising models.
Practical considerations on the improvement of the quality of random
numbers are discussed as well.

\end{abstract}
\begin{keyword}
Random Number Generation, Shift Registers, Lagged Fibonacci,
Cluster Algorithm, Random Walk, Ising Model
\end{keyword}
\end{frontmatter}

\section{Introduction}

\typeout{SET RUN AUTHOR to \@runauthor}

The large-scale simulations in statistical physics use some deterministic
procedures to generate a sequence of uniformly distributed (pseudo)random
numbers.
 It is possible  to generate  $10^8$ numbers per second and $10^{15}$
numbers in 100 days on the best processors. The widely known linear
congruential methods cannot be used in a
such simulations on $32$-bit computers because of the short period
$T<2^{32}\approx 10^9$ of the generated sequence \cite{Knuth2}.
The next most used class of generators are based on the linear recursion
\begin{equation}
X_n=(X_{n-q}\;\; {\sf OP}\;\; X_{n-p})\;\; {\rm mod}\; m,
\label{gen-form}
\end{equation}
where $m$ is the word length and $p$ and $q$ are feedback taps $p>q$.
Here {\sf OP} stands for the exlusive-or operation for shift registers,
$+$ operation for the lagged Fibonacci (LF) and $-$ operation for
the substract-with-carry (SWC) recipes.
In the latter case one has to add
unity if the result of the previous substraction was negative.
The correlations in 2-tap generators were found to be crucial in
large-scale
simulations of Ising model, self-avoiding random walks
\cite{Grassberger} , percolation \cite{ziff-4tap} and cluster algorithms
\cite{FLW} (see, also, \cite{many}).
It was shown recently by Heringa, Bl\"ote and author that these deviations
can be rigorously estimated numerically and even analytically\cite{ljh}.
Using this idea, Bl\"ote and author discovered the scaling of
systematic deviations for Ising model\cite{lh}.

The method gives us some rigorous and quantitative way to estimate the
possible level of deviations in the results of simulations. So, we propose
some practical definition of goodness of a concrete recipe for
generating random numbers. This definition is based on the
concept of observability:
the recipe should be considered as a good one if it passes all known
tests using the sequence of the pseudo-random numbers generated in a
couple of month on the nowadays computers. So, for today we need those
recipes which do not produce observable deviations when $10^{15}$ random
numbers are used.

\section{Random walk test}

The test is based on the one-dimensional directed random walk \cite{ljh}:
a walker starts at some site of an one-dimensional lattice and, at
discrete times $i$, either he takes a step in a  fixed  direction with a
probability $\mu_i$ or he stops with a probability $\nu_i\equiv 1-\mu_i$.
In the case of unbiased probabilities $\mu_i=\mu$ and the probability of a
walk with length $n$ is then $P(n)=\mu^{n-1}\; (1-\mu )$.
We could apply some particular  recipe of random number generation to
calculation of the
probabilities $\mu_i$. Clearly, they would differ from $\mu$ because
of the correlations in the recipe (\ref{gen-form}). Therefore, it was
proposed in \cite{ljh,lh} to calculate the relative deviations
\begin{equation}
\delta P(n)\equiv\frac{P^*(n)}{P(n)}-1
\end{equation}
of the probability of the walk
$P^*(n)=\nu^*_i \, \prod_{i=1}^n (\mu^*_i)^{i-1}$ from the unbiased
probability $P(n)$.

Some results for SR anf LF generators from \cite{ljh} and for SWC one
from \cite{lp}
are presented in the Table~\ref{t1}. The first interesting fact is that
the lagged Fibonacci
and substract-with-carry leads to the same deviations up to the order of
$2^{-m}$, which is usually of the order of $10^{-9}$. Next, the deviations
for shift registers are of the same nature and and of the same order as
for LF and SWC.
Both these statements contradict to the widespread belief that LF
and SWC produce "better" randomness.

\begin{table*}
\caption{Deviations $\delta P(n)$ of the probability of walk length $n$
due to the correlations in the random number sequences. Corrections of the
order of $2^{-m}$ are not included in results for LF and SWC. }
\vspace{2mm}
\label{t1}
\begin{center}
\begin{tabular}{l|c|c}
\hline
length $n$ & 2-tap SR & 2-tap LF $=$ SWC \\
\hline
$p$ & $\frac{1-\mu}{\mu} $ & $\frac{1-2\mu}{2\mu}$ \\ \hline
$p+1$ & $\frac{(2\mu-1)^2}{\mu^4}-1$ & $\frac{(3\mu-1)^2}{4\mu^4}-1$ \\
\hline
\end{tabular}
\end{center}
\end{table*}

\section{Scaling of the deviations}

We performed Wolff simulations of the 2D Ising model at criticality,
using SR with feed-back positions $(p,q)$=(36,11), (89,38),
(127,64) and (250,103) for a variety of the lattice sizes $L$ \cite{lh}.
We found that the data could be collapsed on a single curve with the
appropriate rescaling of deviations, say, of the energy
$\delta \tilde{E} \equiv  p^{0.88} \delta E$
and the system size
$\tilde{L} \equiv p^{-0.43(5)} L$.
This enables us to bound deviations of energy $E$
\begin{equation}
\delta E \,\hbox{\lower0.6ex\hbox{$\sim$}\llap{\raise0.6ex\hbox{$<$}}} \,
0.3\; L^{-0.84}\; p^{-0.52},         \label{dE}
\end{equation}
of the specific heat $C$
\begin{equation}
-\delta C \,\hbox{\lower0.6ex\hbox{$\sim$}\llap{\raise0.6ex\hbox{$<$}}} \,
0.85 \; L^{-0.21}\; p^{-0.42},   \label{dC}
\end{equation}
and of the universal ratio $Q=\langle m^2 \rangle^2/\langle m^4 \rangle$,
where $m$ is the magnetisation,
\begin{equation}
\delta Q \,\hbox{\lower0.6ex\hbox{$\sim$}\llap{\raise0.6ex\hbox{$<$}}} \,
0.244 \, L^{-0.45}\; p^{-0.41}.     \label{dQ}
\end{equation}

The analysis has been performed also for 3D Ising model, and it yields
\begin{equation}
-\delta C\;  \approx \; L^{-0.24}\; p^{-0.19}   \label{dC3}
\end{equation}
and
\begin{equation}
\delta Q \; \approx \;  L^{-1.15}\; p^{0.09}.     \label{dQ3}
\end{equation}

These estimations are very useful in the planning of computer experiments
\cite{spp3}.
Their knowledge is equivalent to the knowledge of the accuracy of the
experimental equipment.

\section{Consclusion}

Using the method developed in \cite{ljh} we found some support for
the previously proposed modifications of the 2-tapped methods.
Correlations would be sharply diminished by one of the following
tricks:
\begin{enumerate}
\item[i)] concept of luxuries \cite{lusher,lp}, i.e., generate $r>p$
random numbers but use only $p$ of them;
\item[ii)] by decimation of the random number sequence, or, by
using a combination of the two or more generators \cite{At,lh,spp3};
\item[iii)] by using four-tap generators \cite{At,ziff-4tap}.
\end{enumerate}

Our results are shown to be important for the simulations of
Ising model, the percolation problem, for random walks and cluster
algorithms.

So, instead of the qualifications "bad" or "good" we propose to apply to
random number generators a more detailed description of their features
\cite{lp}, for example, that no deviations would be observed in the
simulation of the above mentioned problems when up to $10^{15}$ random
numbers are used.

Discussions and collaboration with H. Bl\"ote, J. Herringa and P. Butera
are kindly acknowledged.


\begin{thebibliography}{999}
\bibitem{many} references given in \cite{ljh,lh,lp,ziff-4tap} are cover
               the recent history of the subject
\end{thebibliography}
\end{document}